\documentclass[aps,prb,twocolumn,showpacs,floatfix,longbibliography,superscriptaddress]{revtex4-2}
\usepackage{graphicx,amsfonts,amssymb,amsmath}
\usepackage{textcomp}
\usepackage{esint}
\usepackage[colorlinks,linktocpage,citecolor=blue,linkcolor=red,urlcolor=blue]{hyperref}
\usepackage[utf8]{inputenc}
\usepackage[english]{babel}
\usepackage{afterpage}
\usepackage{bbold}
\usepackage{soul}

\input cyracc.def

\allowdisplaybreaks

\def\kin{k_{\rm in}}
\def\epj{E_J} 
\def\epjk{E_{J,k}}  
\def\tepj{\tilde E_J} 
\def\tepjk{\tilde E_{J,k}} 
\def\epc{E_C} 
\def\epck{E_{C,k}}   
\def\tepc{\tilde  E_C}
\def\tepck{\tilde E_{C,k}} 
\def\bl{\bar l_2^3}
\def\blp{\bar l_0^{2,1}{}'} 

\begin{document}
	

\def\rhoeq{\hat\rho_{\rm eq}}

\newcommand{\marge}[1]{\marginpar{\scriptsize #1}}
\newcommand{\remarque}[1]{\marginpar{\scriptsize Remarque}{\it [#1]}}
\newcommand{\new}[1]{{\bf #1}}
\newcommand{\red}[1]{\textcolor{red}{#1}}
\newcommand{\blue}[1]{\textcolor{blue}{#1}}
\newlength{\textlarg}
\newcommand{\redbar}[1]{\textcolor{red}{\st{#1}}} 
\newcommand{\bluebar}[1]{\textcolor{blue}{\st{#1}}} 

\newcommand{\normord}[1]{:\mathrel{#1}:}

\newcommand{\beq}{\begin{equation}}
\newcommand{\eeq}{\end{equation}}
\newcommand{\bfig}{\begin{figure}}
\newcommand{\efig}{\end{figure}}
\newcommand{\bline}{\begin{multline}}
\newcommand{\eline}{\end{multline}}
\newcommand{\bremark}{\begin{quotation} \noindent \small }
\newcommand{\eremark}{\end{quotation}}
\newcommand{\llbrace}{\left\lbrace}  
\newcommand{\rrbrace}{\right\rbrace}
\newcommand{\lbraket}{\left[}
\newcommand{\rbraket}{\right]}
\newcommand{\llangle}{\left\langle}
\newcommand{\rrangle}{\right\rangle} 

\newcommand{\Tr}{{\rm Tr}} 
\newcommand{\tr}{{\rm tr}} 
\newcommand{\sgn}{\,{\rm sgn}} 
\newcommand{\mean}[1]{\langle #1 \rangle}
\newcommand{\commu}[2]{[#1,#2]} 
\newcommand{\bra}[1]{\langle#1|}
\newcommand{\ket}[1]{|#1\rangle}
\newcommand{\braket}[2]{\langle #1|#2\rangle}
\newcommand{\ketbra}[2]{|#1\rangle\langle#2|}
\newcommand{\dbraket}[3]{\langle #1|#2|#3\rangle}
\newcommand{\tens}[1]{\overleftrightarrow{#1}}  
\newcommand{\vac}{|{\rm vac}\rangle} 
\newcommand{\bravac}{\langle{\rm vac}|}
\newcommand{\const}{{\rm const}} 
\newcommand{\unif}{{\rm unif.}} 
\newcommand{\atanh}{\,{\rm atanh}}
\newcommand{\cotanh}{\,{\rm cotanh}}

\newcommand{\ie}{i.e.\xspace}
\newcommand{\iet}{i.e.}
\newcommand{\eg}{e.g.\xspace}
\newcommand{\cc}{{\rm c.c.}} 
\newcommand{\hc}{{\rm h.c.}} 
\newcommand{\etal}{{\it et al. }}
\newcommand\eme{$^{\mbox{\footnotesize ème}}$\xspace}

\newcommand{\jhatbf}{\hat {\textbf \jold}} 
\newcommand{\Jhatbf}{\hat {\textbf \J}} 
\newcommand{\jhat}{\hat {\jmath}} 
\newcommand{\Jhat}{\hat {J}} 
\newcommand{\jbf}{\textbf j}
\newcommand{\Jbf}{\textbf J}

\def\chibf{\boldsymbol{\chi}}
\def\down{\downarrow}
\def\eps{\epsilon}
\def\gam{\gamma} 
\def\alphabf{\boldsymbol{\alpha}}
\def\phibf{\boldsymbol{\phi}}
\def\varphibf{\boldsymbol{\varphi}}
\def\varphibfs{\boldsymbol{\varphi}_<}
\def\varphibfl{\boldsymbol{\varphi}_>}
\def\varphis{\varphi_{<}}
\def\varphil{\varphi_{>}}
\def\psibf{\boldsymbol{\psi}}
\def\thetabf{\boldsymbol{\theta}}
\def\Ome{\Omega}
\def\omeD{{\omega_D}} 
\def\bfOme{\boldsymbol{\Omega}} 
\def\Omebf{\boldsymbol{\Omega}} 
\def\lamb{\lambda}
\def\Lamb{\Lambda}
\def\sig{\sigma}
\def\Sig{\Sigma}
\def\sigp{{\sigma'}} 
\def\bfsig{\boldsymbol{\sigma}} 
\def\sigbf{\boldsymbol{\sigma}} 
\def\bfSig{\boldsymbol{\Sigma}} 
\def\The{\Theta} 
\def\up{\uparrow}

\def\epsk{\epsilon_{\bf k}} 
\def\xik{\xi_{\bf k}} 
\def\txik{\tilde\xi_{\bf k}} 
\def\xip{\xi_{\bf p}} 
\def\epsq{\epsilon_{\bf q}} 
\def\xiq{\xi_{\bf q}} 
\def\xikq{\xi_{{\bf k}+{\bf q}}} 
\def\Ek{E_{\bf k}} 
\def\Ep{E_{\bf p}}
\def\Eq{E_{\bf q}}
\def\Heff{\hat H_{\rm eff}}
\def\Hem{\hat H_{\rm em}}
\def\Hint{\hat H_{\rm int}}
\def\Hloc{\hat H_{\rm loc}}
\def\HMF{\hat H_{\rm MF}}
\def\HLL{\hat H_{\rm LL}}
\def\Sem{S_{\rm em}}
\def\SMF{S_{\rm MF}} 
\def\SHF{S_{\rm HF}} 
\def\SRPA{S_{\rm RPA}} 
\def\Sint{S_{\rm int}} 
\def\Sloc{S_{\rm loc}}
\def\TN{T_{\rm N}} 
\def\TNHF{T^{\rm HF}_{\rm N}} 
\def\Zloc{Z_{\rm loc}} 
\def\ZMF{Z_{\rm MF}} 
\def\ZHF{Z_{\rm HF}} 
\def\ZRPA{Z_{\rm RPA}} 
\def\RPA{{\rm RPA}}
\def\loc{{\rm loc}} 
\def\pp{{\rm pp}}
\def\ph{{\rm ph}} 
\def\ch{{\rm ch}}
\def\sp{{\rm sp}} 
\def\qtf{q_{\rm TF}}
\def\epstf{\eps^{}_{\rm TF}} 
\def\epsrpa{\eps^{}_{\rm RPA}} 
\def\chinnzpp{\chi_{nn}^{0}{}\!\!\!''}

\def\half{\frac{1}{2}}
\def\dhalf{\dfrac{1}{2}}
\def\third{\frac{1}{3}} 
\def\quarter{\frac{1}{4}}

\def\qr{{\bf q}\cdot{\bf r}}
\def\wt{\omega t} 

\def\a{{\bf a}}
\def\b{{\bf b}}
\newcommand{\cv}{{\bf c}} 
\def\e{{\bf e}}
\def\f{{\bf f}}
\def\g{{\bf g}}
\def\h{{\bf h}}
\def\jold{\char"11}
\def\j{{\bf j}}
\def\k{{\bf k}}
\def\l{{\bf l}}
\def\ellbf{\bm{\ell}} 
\def\m{{\bf m}}
\def\n{{\bf n}} 
\def\p{{\bf p}} 
\def\q{{\bf q}}
\def\r{{\bf r}}
\def\t{{\bf t}}
\def\u{{\bf u}}
\newcommand{\vv}{{\bf v}}
\def\x{{\bf x}}
\def\y{{\bf y}} 
\def\z{{\bf z}} 
\def\A{{\bf A}}
\def\B{{\bf B}}
\def\D{{\bf D}} 
\def\E{{\bf E}} 
\def\F{{\bf F}} 
\def\H{{\bf H}}  
\def\J{{\bf J}}
\def\K{{\bf K}} 

\def\G{{\bf G}}
\def\L{{\bf L}}
\def\M{{\bf M}}  
\def\O{{\bf O}} 
\def\P{{\bf P}} 
\def\Q{{\bf Q}} 
\def\R{{\bf R}}
\def\S{{\bf S}}
\def\U{{\bf U}} 
\def\V{{\bf V}} 
\def\X{{\bf X}} 
\def\Y{{\bf Y}} 
\def\epsbf{\boldsymbol{\epsilon}}
\def\betabf{\boldsymbol{\beta}}
\def\deltabf{\boldsymbol{\delta}}
\def\mubf{\boldsymbol{\mu}}
\def\nablabf{\boldsymbol{\nabla}}
\def\rhobf{\boldsymbol{\rho}}
\def\sigmabf{\boldsymbol{\sigma}} 
\def\Pibf{\boldsymbol{\Pi}}
\def\pibf{\boldsymbol{\pi}}

\def\para{\parallel}
\def\kpara{{k_\parallel}}
\def\kperp{{k_\perp}} 
\def\kperpp{{k_\perp'}} 
\def\qperp{{q_\perp}} 
\def\tperp{{t_\perp}} 

\def\w{\omega}
\def\wn{\omega_n}
\def\wm{\omega_m}
\def\wnu{\omega_\nu}
\def\wp{\omega_p} 
\def\dmu{{\partial_\mu}}
\def\dnu{{\partial_\nu}}
\def\dl{{\partial_l}}  
\def\dt{\partial_t} 
\def\tdt{\tilde\partial_t}
\def\dk{\partial_k}
\def\tdk{\tilde\partial_k}
\def\dx{\partial_x}
\def\dy{\partial_y} 
\def\dw{\partial_{\w}}
\def\dtau{{\partial_\tau}}  
\def\det{{\rm det}} 
\def\Pf{{\rm Pf}}
\def\diag{{\rm diag}}

\def\dsum{\displaystyle \sum}
\def\dint{\displaystyle \int} 
\def\intt{\int_{-\infty}^\infty dt} 
\def\inttp{\int_{-\infty}^\infty dt'} 
\def\intk{\int_{\bf k}} 
\def\intkd{\int \frac{d^dk}{(2\pi)^d}}
\def\intq{\int_{\bf q}} 
\def\intr{\int d^dr}  
\def\dintr{\displaystyle \int d^dr} 
\def\intrp{\int d^dr'}
\def\dinttau{\displaystyle \int_0^\beta d\tau}
\def\dinttaup{\displaystyle \int_0^\beta d\tau'}
\def\inttau{\int_0^\beta d\tau}
\def\inttaup{\int_0^\beta d\tau'}
\def\intx{\int d^{d+1}x} 
\def\inttaur{\int_0^\beta d\tau \int d^dr}
\def\intinf{\int_{-\infty}^\infty}
\def\dinttaur{\displaystyle \int_0^\beta d\tau \int d^dr}
\def\dintinf{\displaystyle \int_{-\infty}^\infty}
\def\intw{\int_{-\infty}^\infty \frac{d\w}{2\pi}}
\def\sumr{\sum_{\bf r}} 

\def\calA{{\cal A}}
\def\calAbf{\bm{{\cal A}}}
\def\calB{{\cal B}} 
\def\calC{{\cal C}} 
\def\dt{\partial_t}
\def\calD{{\cal D}}
\def\calE{{\cal E}}
\def\calF{{\cal F}} 
\def\calFbf{\bm{{\cal F}}}
\def\calG{{\cal G}}
\def\calH{{\cal H}}
\def\calI{{\cal I}}
\def\calJ{{\cal J}}
\def\calK{{\cal K}}
\def\calL{{\cal L}} 
\def\calM{{\cal M}} 
\def\calN{{\cal N}}
\def\calO{{\cal O}}
\def\calP{{\cal P}}  
\def\calR{{\cal R}} 
\def\calS{{\cal S}}
\def\calT{{\cal T}}
\def\calU{{\cal U}}
\def\calV{{\cal V}}
\def\calX{{\cal X}} 
\def\calY{{\cal Y}} 
\def\calW{{\cal W}} 
\def\calZ{{\cal Z}}

\def\tT{{\tilde T}}
\def\talpha{{\tilde\alpha}}
\def\tbeta{{\tilde\beta}}
\def\tchi{{\tilde\chi}}
\def\tdelta{{\tilde\delta}}
\def\tDelta{{\tilde\Delta}}
\def\teta{{\tilde\eta}} 
\def\tlamb{{\tilde\lambda}}
\def\tmu{{\tilde\mu}}
\def\tphibf{{\tilde\phibf}}
\def\trho{{\tilde\rho}}
\def\tvarphibf{{\tilde\varphibf}} 
\def\tw{{\tilde\omega}}
\def\twn{{\tilde\omega_n}}
\def\twnu{{\tilde\omega_\nu}}

\def\asinh{{\rm asinh}} 
\def\Tbkt{T_{\rm BKT}}
	
\graphicspath{{./figures_submit/}}
	
\title{Nature of the Schmid transition in a resistively shunted Josephson junction}
	
\author{Romain Daviet} 
\affiliation{Institut f\"ur Theoretische Physik, Universit\"at zu K\"oln, D-50937 Cologne, Germany}
\author{Nicolas Dupuis}
\affiliation{Sorbonne Universit\'e, CNRS, Laboratoire de Physique Th\'eorique de la Mati\`ere Condens\'ee, LPTMC, F-75005 Paris, France}
	
\date{December 13, 2023} 
	
\begin{abstract}
We study the phase diagram of a resistively shunted Josephson junction (RSJJ) in the framework of the boundary sine-Gordon model. Using the non-perturbative functional renormalization group (FRG) we find that the transition is not controlled by a single fixed point but by a line of fixed points, and compute the continuously varying critical exponent $\nu$. We argue that the conductance also varies continuously along the transition line. In contrast to the traditional phase diagram of the RSJJ ---an insulating ground state when the shunt resistance $R$ is larger than $R_q=h/(2e)^2$ and a superconducting one when $R<R_q$--- the FRG predicts the transition line in the plane $(\alpha,\epj/\epc)$ to bend in the region $\alpha=R_q/R<1$ but we cannot discard the possibility of a vertical line at $\alpha=1$ ($\epj$ and $\epc$ denote the Josephson and charging energies of the junction, respectively). Our results regarding the phase diagram and the nature of the transition are compared with Monte Carlo simulations and numerical renormalization group results. 
\end{abstract}
\pacs{} 
	
\maketitle
 
\tableofcontents
	
\section{Introduction.} 
	
Although the resistively shunted Josephson junction (RSJJ)~\cite{Schoen90} is one of the best-studied examples of dissipative quantum systems~\cite{Caldeira83a}, its basic properties are still a matter of debate. Since the seminal work of Schmid and Bulgadaev (SB) on the quantum Brownian particle in a periodic potential (a problem that can be mapped on the RSJJ)~\cite{Schmid83,Bulgadaev84}, the conventional view is that the junction is insulating when the shunt resistance $R$ is larger than the quantum of resistance $R_q=h/q^2$ ($q=2e$ is the Cooper pair charge) and superconducting when $R<R_q$~\cite{Guinea85,Fisher85a,Aslangul87,Schoen90} but for a long time there has been little experimental evidence~\cite{Yagi97,Penttilae99,Penttilae01,Kuzmin91}. In a recent experiment no sign of the expected insulating phase was observed and the very existence of the dissipative quantum phase transition at $R=R_q$ has been questioned~\cite{Murani20,Hakonen21,*Murani21}, whereas the observation of the Schmid transition has been reported in another experiment~\cite{Kuzmin23}. On the other hand, using both the numerical renormalization group (NRG) and the functional renormalization group (FRG), it has been shown that when $\alpha=R_q/R<1$ an insulating-superconducting transition can be induced by varying the ratio $\epj/\epc$ between the Josephson coupling energy and the charging energy~\cite{Masuki22,Yokota23}, but these conclusions have been questioned~\cite{Sepulcre22,*Masuki22a}.  
	
\begin{figure}
		\centerline{\includegraphics[width=3.5cm]{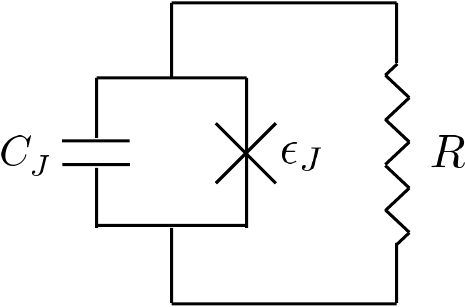}}
		\caption{Resistively (and capacitively) shunted Josephson junction (RSJJ). The capacitance $C_J$ determines the charging energy $E_C=(2e)^2/2C_J$ of the junction while the transparency of the tunnel barrier and the superconducting gap set the Josephson coupling energy $E_J$.}
		\label{fig_RSJJ} 
\end{figure}

In this paper, we reconsider the superconductor-insulator transition in a RSJJ, as shown in Fig.~\ref{fig_RSJJ}, where the Josephson junction is shunted by both a resistance and a capacitance~\cite{not22}. The RSJJ can be described in the framework of the boundary sine-Gordon model originally studied by SB and defined by the Euclidean (imaginary-time) action~\cite{not16a}  
\begin{align} 
		S[\varphi] ={}& \half \sum_{\wn} \left( \frac{\alpha}{2\pi} |\wn| + \frac{\wn^2}{2E_C} \right)  |\varphi(i\wn)|^2 \nonumber\\ &  
		- E_J \inttau \cos \varphi(\tau) , 
		\label{action}
\end{align} 
where $E_J$ is the Josephson coupling energy, $E_C=q^2/2C_J$ the charging energy and $C_J$ the capacitance of the junction. The field $\varphi$, which stands for the superconducting order parameter phase difference across the junction, is a noncompact variable which satisfies periodic boundary conditions in imaginary time, $\varphi(0)=\varphi(\beta)$, and $\wn=2n\pi T$ ($n$ integer) is a bosonic Matsubara frequency. $\beta=1/T$ is the inverse temperature and we consider only the zero-temperature limit $\beta\to\infty$ (we set $\hbar=k_B=1$ throughout the paper). We assume a ultraviolet (UV) frequency cutoff $W$. The action~(\ref{action}) also describes a quantum Brownian particle in one dimension with coordinate $\varphi$ and mass $m=1/2\epc$ moving in a periodic potential ($\eta=\alpha/2\pi$ is then the friction coefficient in the classical limit)~\cite{Schmid83,Bulgadaev84,Fisher85a}, and a Luttinger liquid in presence of an impurity (with $K=1/\alpha$ the Luttinger parameter) or a weak link (with $K=\alpha$)~\cite{Kane92a,Kane92,not12}.
	
We study the boundary sine-Gordon model~(\ref{action}) in the framework of the nonperturbative FRG~\cite{Berges02,Delamotte12,Dupuis_review}, an approach which has proven very successful for the $(1+1)$-dimensional sine-Gordon model~\cite{Daviet19,Jentsch22}. We find that the transition is not controlled by a single fixed point but by a line of fixed points, and compute the continuously varying critical exponent $\nu$ associated with the relevant direction about the fixed point. These results qualitatively agree with the Monte Carlo simulations of Werner and Troyer (WT) who showed that the correlation function $\chi(\tau)=\mean{e^{iq\varphi(\tau)} e^{-iq\varphi(0)}}$ ($|q|\leq 1/2$) exhibits continuously varying critical exponents along the transition line~\cite{Werner05,Lukyanov07}. Recent NRG calculations also imply that the transition line is a line of fixed points~\cite{Masuki22}. Although no precise calculations are performed, we argue that the conductance varies continuously along the transition line. 

Unlike the traditional phase diagram of the RSJJ, where the transition between the insulating and superconducting phases is located at $\alpha=1$, the FRG predicts the transition line in the plane $(\alpha,\epj/\epc)$ to bend in the region $\alpha=R_q/R<1$. The FRG is however not reliable in the limit $\alpha\to 0$ since it predicts a phase transition at a finite value of $E_J/E_C$ while in the absence of dissipation the ground state of the model~(\ref{action}) is known to be insulating. This leads us to propose two possible scenarios for the Schmid transition. In the first one, the transition line is vertical and located at $\alpha=1$, as in the traditional phase diagram and in agreement with the Monte Carlo simulations of WT. In the second one, the transition line bends in the region $\alpha<1$, as in the FRG and NRG calculations~\cite{Masuki22,Yokota23}, but eventually the critical value of $E_J/E_C$ diverges as $\alpha\to 0$. 

On the other hand we compute the phase mobility $\mu(\w)$ (i.e. the mobility of the Brownian particle) related to the admittance $Y(\w)=q^2/\mu(\w)$ of the RSJJ. The dc mobility $\mu=\lim_{\w\to 0}\mu(\w)$ vanishes in the superconducting phase, and  is equal to the mobility $\mu_0=2\pi/\alpha=1/\eta$ of the free particle in the insulating phase. The frequency dependence of the mobility in the superconducting phase is correctly obtained only for $\alpha\gtrsim 2$; when $1\leq\alpha\lesssim 2$, the FRG fails to capture the instantons connecting neighboring minima of the periodic potential. This does not prevent us to obtain the low-frequency behavior of the RSJJ; in the insulating phase $Y(\w\to 0)=1/R$ whereas $Y(\w)$ is purely inductive at low frequencies in the superconducting phase. However the effective inductance is not determined by the ``coherence'', i.e. the expectation value $\mean{\cos\varphi}$, which in fact remains nonzero in the insulating phase. We compare our results for $\mean{\cos\varphi}$ and $\mean{\cos(\varphi/2)}$ with results obtained from integrability methods and Monte Carlo simulations~\cite{Fateev97,Lukyanov07}.

\section{FRG formalism.}

Following the standard strategy of the nonperturbative functional renormalization group (FRG) we add to the action an infrared regulator term~\cite{Berges02,Delamotte12,Dupuis_review},
\beq 
\Delta S_k[\varphi] = \half \sum_{\wn} R_k(i\wn) \varphi(-i\wn) \varphi(i\wn) , 
\eeq 
which suppresses fluctuation modes whose frequency is smaller than the (running) frequency $k$, i.e. $|\wn|\lesssim k$, but leaves unaffected those with $|\wn|\gtrsim k$. The cutoff function is written in the form 
\beq 
R_k(i\wn) = \frac{\alpha}{2\pi} |\wn| r\left( \frac{|\wn|}{k} \right) + Z_{2,k} \wn^2 r\left( \frac{\wn^2}{k^2} \right) , 
\label{Rdef} 
\eeq 
where $Z_{2,k}=\frac{1}{2\pi}\int_0^{2\pi} d\phi\, Z_{2,k}(\phi)$ is the field average of the function $Z_{2,k}(\phi)$ defined in Eq.~(\ref{Gamma1}) and $r(x)=4(1-x)^2/x\,\Theta(1-x)$. Including the second term in~(\ref{Rdef}) allows us to consider the limit $\alpha\to 0$. The partition function 
\beq 
\calZ_k[J] = \int \calD[\varphi] \, e^{-S[\varphi] - \Delta S_k[\varphi] + \inttau\, J\varphi } 
\eeq 
thus becomes $k$ dependent. The expectation value of the field is given by 
\beq 
\phi(\tau) = \frac{\delta \ln \calZ_k[J]}{\delta J(\tau)} = \mean{\varphi(\tau)} . 
\eeq 
The scale-dependent effective action 
\beq 
\Gamma_k[\phi] = - \ln \calZ_k[J] + \inttau\, J\phi - \Delta S_k[\phi] 
\label{Gammadef}
\eeq
is defined as a slightly modified Legendre transform which includes the subtraction of $\Delta S_k[\phi]$. Assuming that for $k=\kin$ the fluctuations are completely frozen by the term $\Delta S_{\kin}$, which is the case when $\kin\gg\min(W,\epc)$, $\Gamma_{\kin}[\phi]=S[\phi]$. On the other hand, the effective action of the original model~(\ref{action}) is given by $\Gamma_{k=0}[\phi]$ since $R_{k=0}$ vanishes. The nonperturbative FRG approach aims at determining $\Gamma_{k=0}$ from $\Gamma_{\kin}$ using Wetterich’s equation~\cite{Wetterich93,Ellwanger94,Morris94} 
\beq 
\dt \Gamma_k[\phi] = \half \Tr\Bigl\{ \dt R_k \bigl( \Gamma_k^{(2)}[\phi] + R_k\bigr)^{-1} \Bigr\} ,
\label{eqwet} 
\eeq
where $\Gamma_k^{(2)}$ is the second-order functional derivative of $\Gamma_k$ and $t=\ln(k/\kin)$ a (negative) RG “time”. 

\subsection{FE$_2$ expansion}

In the frequency expansion to second order (FE$_2$)~\cite{not9}, the scale-dependent effective action is approximated by 
\begin{align}
	\Gamma_k[\phi] ={}& \frac{Z_1}{2} \sum_{\wn} |\wn| \phi(-i\wn) \phi(i\wn) \nonumber\\& 
	+ \inttau \left[ \half Z_{2,k}(\phi) (\dtau\phi)^2 + U_k(\phi) \right] 
	\label{Gamma1} 
\end{align} 
with initial conditions 
\beq 
Z_1 = \frac{\alpha}{2\pi} , \quad Z_{2,\kin}(\phi) = \frac{1}{2E_C} , \quad U_{\kin}(\phi) = -E_J \cos\phi . 
\eeq 
The effective potential $U_k(\phi)$ is given by the effective action when the field $\phi(\tau)\equiv \phi$ is time independent: $\Gamma_k[\phi]=\beta U_k(\phi)$. We anticipate the fact that $Z_1$ is not renormalized and therefore remains equal to its initial value. The zeroth-order harmonic of $Z_{2,k}(\phi)$ and the first-order harmonic of $U_k(\phi)$ can be seen as renormalized values of the coupling constants $1/2\epc$ and $\epj$, respectively, but the flow equation generates higher-order harmonics.  

In practice, one introduces the dimensionless quantities
\beq 
\tilde U_k(\phi) = \frac{U_k(\phi)}{k} , \qquad \tilde Z_{2,k}(\phi) = k Z_{2,k}(\phi) ,
\label{dimless} 
\eeq  
which ensure that the zero-temperature quantum phase transition between the superconducting and insulating phases corresponds to a fixed point of the flow equations. The latter, which are given in Appendix~\ref{app_floweq}, must be solved numerically.

\subsection{Mobility and admittance} 

In addition to the effective potential $U_k(\phi)$, whose RG flow indicates whether the RSJJ is insulating or superconducting, a fundamental quantity is the mobility $\mu(i\wn)=|\wn|G(i\wn)$ of the quantum Brownian particle, i.e. its average (long-time) velocity when it is subjected to an external force. Since the phase propagator $G_k(i\wn,\phi)=1/\Gamma^{(2)}_k(i\wn,\phi)$ is given by the inverse of the two-point vertex, whose most general expression reads $\Gamma_k^{(2)}(i\wn,\phi)=Z_1|\wn|+\Delta_k(i\wn,\phi)+U''_k(\phi)$ (for a time-independent field), the scale-dependent mobility reads 
\beq 
\mu_k(i\wn) = \frac{|\wn|}{\frac{\alpha}{2\pi}|\wn| + \Delta_k(i\wn,0) + U''_k(0)} . 
\label{mob2} 
\eeq 
We consider the vanishing field configuration $\phi=0$ since this corresponds to the minimum of the effective potential for any $k>0$. In the FE$_2$, the self-energy $\Delta_k$ is approximated by its lowest-order derivative expansion, i.e. $\Delta_k(i\wn,\phi)=Z_{2,k}(\phi)\wn^2$. To obtain the frequency-dependent mobility $\mu(\w)$ in real time, one must first take the limit $k\to 0$ and perform the analytic continuation $i\wn\to \w+i0^+$; the dc mobility is then given by $\mu=\mu(\w\to 0)$. In the FE$_2$, the determination of $\Delta_k(i\wn,\phi)$ is however valid only in the limit $|\wn|\ll k$ ---for the same reason that the derivative expansion in the $\varphi^4$ theory is valid only in the small-momentum limit $|\p|\ll k$~\cite{Dupuis_review}--- and setting $k\to 0$ followed by $\w\to 0$ (or $\wn\to 0$) is not, at least in principle, possible. We shall see how Eq.~(\ref{mob2}) can nevertheless be used to obtain useful information about the mobility. 

When the RSJJ is biased by an infinitesimal external time-dependent current $I(t)$, the mobility determines the induced voltage $V(t)$ across the junction~\cite{Schoen90}, i.e. the admittance 
\beq 
Y(\w) = \frac{I(\w)}{V(\w)} = \frac{q^2}{\mu(\w)} 
\label{Ydef} 
\eeq 
for a sinusoidal current.

\section{The Schmid transition}

In this section, we discuss the phase diagram obtained from the FRG approach and the nature of the transition between the superconducting and insulating phases. We compare our findings with the traditional view of the SB transition and previous numerical results. We also comment on some differences with the FRG results of Masuki {\it et al.}~\cite{Masuki22,Yokota23}.

\begin{figure*}
	\hspace*{-0.5cm}
	\centerline{
		\includegraphics[height=4.8cm]{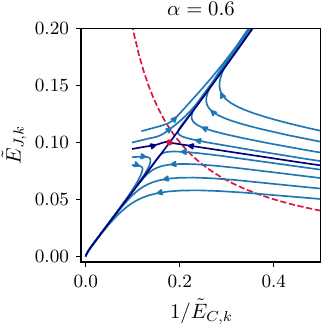} \includegraphics[height=4.8cm]{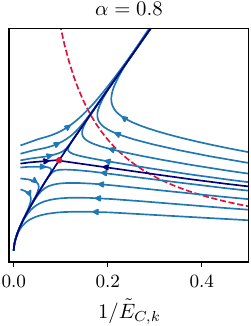} 
		\includegraphics[height=4.8cm]{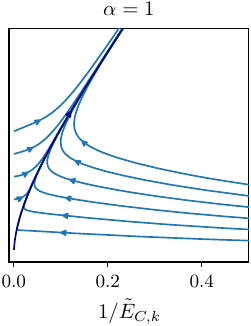} 
		\includegraphics[height=4.8cm]{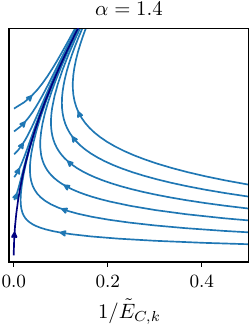}}
	\caption{Flow diagram projected on the $(1/\tepck,\tepjk$) plane, where $1/2\tepck,$ and $\tepjk$ denote the zeroth- and first-order harmonics of $\tilde Z_{2,k}(\phi)$ and $\tilde U_k(\phi)$, respectively ($\alpha=0.6,0.8,1,1.4$ from left to right). The red point shows the fixed point obtained by solving $\dk\tilde U_k(\phi)=\dk\tilde Z_{2,k}(\phi)=0$. The red dashed lines, showing the effective initial conditions of the flow $(1/\tilde E_{C,W}=W/E_C,\tilde E_{J,W}=E_J/W)$ at fixed $E_J/E_C=0.02$ when $W$ is varied \cite{not13}, indicate that a smaller bandwidth $W$ favors the superconducting phase. }  
	\label{fig_flowdiag1} 
\end{figure*}

\begin{figure}
	\centerline{\includegraphics{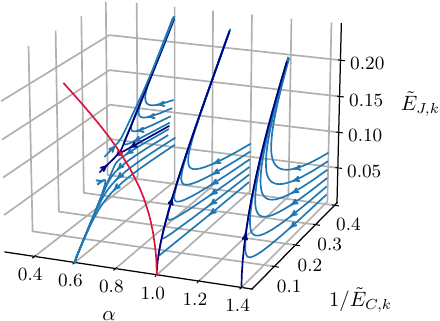}}
	\caption{Flow diagram projected on the $(1/\tepck,\tepjk)$ plane for $\alpha=0.6$, $1$ and $1.4$. The line of critical fixed points for $\alpha<1$ is shown in red.
	}
	\label{fig_flowdiag2} 
\end{figure}  

\subsection{Phase diagram} 

Typical flow trajectories, shown in Figs.~\ref{fig_flowdiag1} and \ref{fig_flowdiag2}, are in agreement with previous results by Masuki {\it et al.}~\cite{Masuki22,Yokota23}. When $\alpha>1$, there is a (repulsive) trivial fixed point $\tilde U'(\phi)=\tilde Z_{2}(\phi)=0$ and all RG trajectories with initial condition $\tilde E_{J,\kin}=\epj/\kin>0$ flow to the strong coupling limit where both the effective potential $\tilde U'(\phi)$ and $\tilde Z_{2}(\phi)$ flow to infinity~\cite{not18}: The system is in the superconducting phase. When $\alpha<1$, in addition to the trivial fixed point $\tilde U'(\phi)=\tilde Z_{2}(\phi)=0$, which is now attractive, we find a critical fixed point $(\tilde U^*{}'(\phi),\tilde Z_2^*(\phi))$, so that the trajectories can flow to either the trivial fixed point or the strong-coupling limit  depending on the initial conditions at scale $k=\kin$. Thus the system can be either superconducting ($\tilde U_k',\tilde Z_{2,k}\to\infty$ for $k\to 0$) or insulating ($\tilde U_k',\tilde Z_{2,k}\to 0$). 

\begin{figure}[b]
	\centerline{\includegraphics{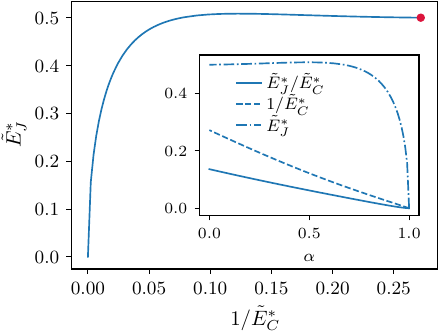}}
\caption{Location of the critical fixed point in the plane $(1/\tepc^*,\tepj^*)$ as $\alpha$ varies. The inset shows $1/\tepc^*$ and $\tepj^*$ vs $\alpha$. Near $\alpha=1$, $1/\tepc^*\sim 1-\alpha$ and $\tepj^*\sim\sqrt{1-\alpha}$. The red point shows the (spurious) fixed point obtained when $\alpha=0$ (see text for a discussion). }
\label{fig_FP} 
\end{figure}

The location of the critical fixed point as $\alpha$ varies can be obtained by solving the equations $\dk \tilde U_k'(\phi)=\dk \tilde Z_{2,k}(\phi)=0$ (Fig.~\ref{fig_FP}).  When $\alpha\to 0$, the action~(\ref{action}) corresponds to the one-dimensional sine-Gordon model and describes a frictionless quantum particle in a one-dimensional periodic potential. Since all quantum states of the particle are extended, the ground state should be insulating in that limit, whatever the value of $\epc$ and $\epj$. The existence of a fixed point at a finite value $\tepj^*/\tepc^*$ when $\alpha=0$ is a known artifact of the FRG-FE$_2$ approach to the one-dimensional sine-Gordon model~\cite{Nandori14}. In the limit $\alpha\to 0$, we expect the fixed point (shown by a red dot in Fig.~\ref{fig_FP}) to move to infinity, i.e. $\tepj^*/\tepc^*\to\infty$~\cite{notad}. 
	
\begin{figure}
	\centerline{\includegraphics{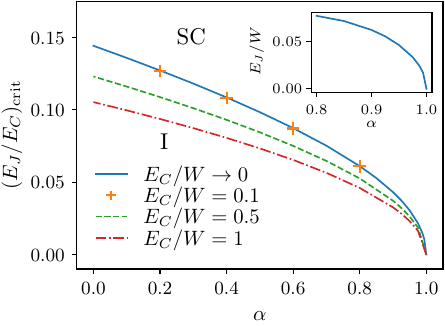}} \vspace{0.5cm}
	\centerline{\includegraphics[width=7.5cm]{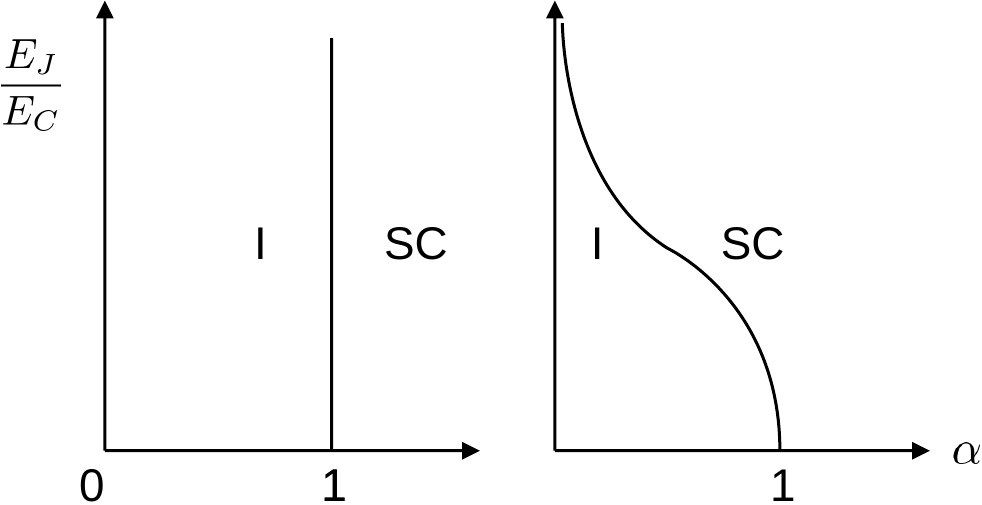}}
	\caption{(Top) Zero-temperature phase diagram of the RSJJ as a function of the bare parameters of the model ($\alpha$, $\epj$, $1/\epc$ and $W$) showing the insulating (I) and superconducting (SC) phases. The transition line depends only on the ratio $\epj/\epc$ when $W\gg\epc$. The inset shows the phase diagram for $W/\epc=0.01$. As explained in the text, the existence of a transition at a finite value $(E_J/E_C)_{\rm crit}$ when $\alpha\to 0$ is an artifact of the FE$_2$. (Bottom) Two possible scenarios for the Schmid transition taking into account the absence of transition in the limit $\alpha\to 0$. 
	}
	\label{fig_phasediag}
\end{figure}

The phase diagram as a function of the bare parameters of the model is shown in Fig.~\ref{fig_phasediag}. We find a transition between an insulating and a superconducting phase in the region $\alpha<1$. In the large bandwidth limit $W/\epc\gg 1$, the transition line depends only on the ratio $\epj/\epc$ and starts with an infinite slope: $(\epj/\epc)_{\rm crit}\sim (1-\alpha)^{0.5}$ when $\alpha\to 1^-$. The transition line then bends towards the region $\alpha<1$, a direct consequence of the existence of the line of critical fixed points shown in Fig.~\ref{fig_FP}. The finite value of $(\epj/\epc)_{\rm crit}$ in the limit $\alpha\to 0$ is due to $\tepj^*$ remaining finite; if $\tepj^*\to\infty$ when $\alpha\to 0$ (as it should be, see the previous discussion) the RSJJ is insulating whatever the value of the Josephson energy $\epj$. Two possible scenarios are shown in Fig.~\ref{fig_phasediag} (bottom panel). The first one is a vertical transition line located at $\alpha=1$, as in the traditional picture of the Schmid transition and in agreement with the Monte Carlo simulations of WT in the large bandwidth limit $W\gg E_C$ who find that the transition occurs at $\alpha=1.00(2)$ at least up to $E_J/E_C\sim 0.5$~\cite{Werner05,not14}. In this scenario, the presence of the spurious fixed point at $\alpha=0$ invalidates the entire transition line obtained in the FE$_2$, the only vestige of the true transition line being the vertical tangent at $\alpha=1$. In the second scenario, the spurious fixed point at $\alpha=0$ invalidates the FE$_2$ result for $\alpha\ll 1$ but the obtained transition line is correct in a finite interval near $\alpha=1$. The NRG results of Masuki {\it et al.}~\cite{Masuki22} also yield a transition line that bends in the region $\alpha<1$ but with two qualitative differences with the FRG results reported here: The transition line exhibits a vanishing slope near $\alpha=1$ and a reentrance of the superconducting phase at small $\alpha$ (which is hard to reconcile with the known phase diagram of the one-dimensional sine-Gordon model ($\alpha=0$), see the discussion above). 

Moving away from the large bandwidth limit, at fixed $\epc$, favors the superconducting phase as shown in Fig.~\ref{fig_flowdiag1}: Decreasing $W$ will always change the initial conditions of the flow so as to make the system superconducting. Thus $(\epj/\epc)_{\rm crit}$ decreases when $W$ is lowered. As shown in the inset of Fig.~\ref{fig_phasediag}, we find that the transition persists in the limit $W/\epc\to 0$ since $(\epj/W)_{\rm crit}$ remains finite. This differs from the conclusion of Ref.~\cite{Masuki22} that a nonzero value of $1/E_C$ is necessary to have a transition for $\alpha<1$. 

\subsection{Critical behavior} 

\subsubsection{Critical exponent $\nu$} 

\begin{figure}
	\centerline{\includegraphics{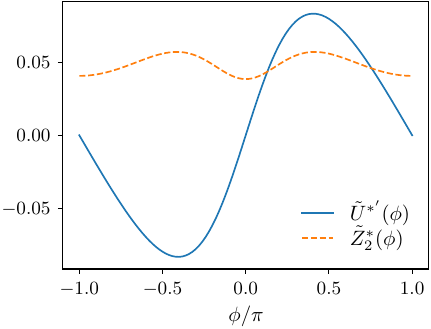}}
	\caption{Fixed-point functions $\tilde U^*{}'(\phi)$ and $\tilde Z^*_2(\phi)$ for $\alpha=0.8$.}
	\label{fig_UZ2tildeFP}
\end{figure}

\begin{figure}[b]
	\centerline{\includegraphics{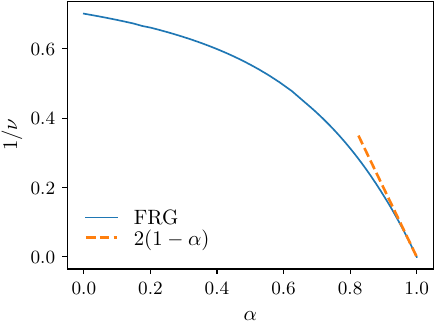}}
	\caption{Critical exponent $\nu$ vs $\alpha$. The dashed (red) line shows the result $1/\nu=2(1-\alpha)$ valid near $\alpha=1$.} 
	\label{fig_nu} 
\end{figure}

The functions $\tilde U^*{}'(\phi)$ and $\tilde Z^*_2(\phi)$ at the fixed point controlling the phase transition are shown in Fig.~\ref{fig_UZ2tildeFP} for $\alpha=0.8$.  The RG eigenvalue $1/\nu$ associated with the relevant perturbation, shown in Fig~\ref{fig_nu}, is determined by linearizing the flow about the fixed point.


Near $\alpha=1$, the critical fixed point is close to the Gaussian fixed point and can be analyzed from perturbation theory. We use the harmonic expansion $\tilde U_k(\phi) = \sum_{n=0}^\infty \tilde u_{n,k} \cos(n\phi)$ and $\tilde Z_{2k}(\phi)=\sum_{n=0}^\infty \tilde z_{n,k} \cos(n\phi)$, and expand the flow equations in powers of $\eps=1-1/\alpha$. When $\eps\to 0$, the flow of $\tilde u_{1,k}$ is initially much slower than the flow of all other coupling constants. After a transient regime, the values of $\tilde u_{n\neq 1,k}$ and $\tilde z_{n,k}$ are determined by $\tilde u_{1,k}$ alone and all RG trajectories collapse on a single line, as shown in Figs.~\ref{fig_flowdiag1} and \ref{fig_flowdiag2} for the trajectories projected on the plane $(\tilde z_0,\tilde u_1)$.  The flow along this line is determined by the beta function
\begin{equation}
	\dt \tilde u_{1,k} = \left( \frac{1}{\alpha} - 1 \right) \tilde u_{1,k}  + \calF \tilde u_{1,k}^3 ,
	\label{pt7}
\end{equation} 
where $\calF$ is a complicated combination of threshold functions (see Appendix~\ref{app_perturbation} for detail). Note that this beta function is not exact (to order $\tilde u^3_{1,k}$) since it relies on the FE$_2$ expansion.

Linearizing the flow equation~(\ref{pt7}) about its nontrivial fixed point $\tilde u_1^*= [(\alpha-1)/\alpha\calF]^{1/2}$ yields the critical exponent $1/\nu=2(1-\alpha)+\calO((1-\alpha)^2)$, regardless of the value of $\calF$. Only the sign of $\calF$ matters (assuming $\calF$ to be nonzero) since it determines whether the fixed point $\tilde u_1^*$ exists when $\alpha<1$ or $\alpha>1$. The fixed point is repulsive if $\alpha<1$ ($1/\nu>0$) and attractive if $\alpha>1$ ($1/\nu<0$), in agreement with the trivial fixed point $\tilde u_1=0$ being attractive if $\alpha<1$ and repulsive if $\alpha>1$. A numerical evaluation yields $\calF<0$ thus indicating that the nontrivial fixed point exists when $\alpha<1$. 

In their Monte Carlo simulations~\cite{Werner05}, WT also observed that the transition is not controlled by a single fixed point, but rather by a line of critical points: By considering the correlation function $\mean{e^{iq\varphi(\tau)} e^{-iq\varphi(0)}}$ ($|q|\leq 1/2$), they obtained continuously varying exponents along the transition line, in full agreement with the FRG analysis, see Sec.~\ref{subsubsec_chiWT}.
The NRG calculations of Masuki {\it et al.} also predict the Schmid transition to be controlled by a line of fixed points~\cite{Masuki22}. 
		
The bending of the transition line in the region $\alpha<1$ clearly simplifies the study of the critical behavior. For a vertical transition line at $\alpha=1$, one would expect the cubic term $\calF$ and all higher-order terms in the beta function~(\ref{pt7}) to vanish, in order to ensure that the transition line is a fixed line. There are various claims in the literature that the cubic term $\calF$ in the beta function vanishes although no detailed calculations seem to have been reported~\cite{Guinea85,Fisher85a}. Interestingly, a nonzero cubic term was found by Bulgadaev in his original paper~\cite{Bulgadaev84}. In the FE$_2$ expansion, the vanishing of the beta function to all orders at $\alpha=1$ is made impossible by the (spurious) fixed point at $\alpha=0$, which is at the origin of the transition line in the region $\alpha<1$.

Note that in the scenario where the line of fixed points and the transition line are vertical, $1/\nu$ vanishes since the beta function $\dt\tilde u_{1,k}=(1/\alpha-1)\tilde u_{1,k}$ is identically zero for $\alpha=1$. The situation is different for the exponent $\gamma$ discussed in the following section.

\begin{figure}
	\centerline{\includegraphics{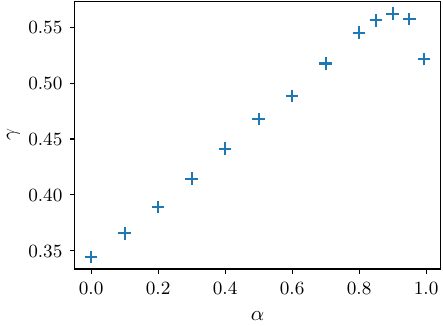}}
	\caption{Critical exponent $\gamma$ associated with the correlation function $\chi(\tau)$ [Eq.~(\ref{chi_def})].}
	\label{fig_gamma_chi} 
\end{figure}

\subsubsection{Correlation function $\mean{e^{\frac{i}{2}\varphi(\tau)} e^{-\frac{i}{2}\varphi(0)}}$} 
\label{subsubsec_chiWT}
	
Following WT~\cite{Werner05,Lukyanov07}, we consider the correlation function 
\beq 
\chi(\tau) = \mean{e^{\frac{i}{2}\varphi(\tau)} e^{-\frac{i}{2}\varphi(0)}} . 
\label{chi_def} 
\eeq 
The computation of its Fourier transform $\chi_k(i\wn)$ for $i\wn=0$ is discussed in Appendix~\ref{app_coherence}. At criticality we find that $\chi_k(i\wn=0)\sim 1/k^{1-\gamma}$ diverges with some exponent $1-\gamma$. By dimensional analysis, we then obtain $\chi_{k=0}(i\wn)\sim 1/|\wn|^{1-\gamma}$ and 
\beq 
\chi_{k=0}(\tau) \sim  \frac{1}{|\tau|^{\gamma}} .
\eeq 
The exponent $\gamma$ is shown in Fig.~\ref{fig_gamma_chi}. The limiting value $\gamma\simeq 0.5$ when $\alpha\to 1$ (i.e. $E_J\to 0$) agrees with the result of Lukyanov and Werner obtained from integrability methods~\cite{Lukyanov07}. We note however that $\gamma$ exhibits a maximum for $\alpha\sim 0.9$ while they obtain a strictly monotonous exponent along the transition line.

The qualitative agreement between our results and those of Lukyanov and Werner (except for the maximum near $\alpha\sim 0.9$) suggests that our determination of the exponent $\gamma$ along the transition line is correct even in the scenario where the transition line is vertical.

\begin{figure}
	\centerline{\includegraphics{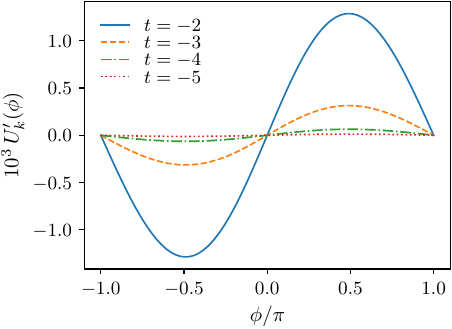}} 
	\centerline{\includegraphics{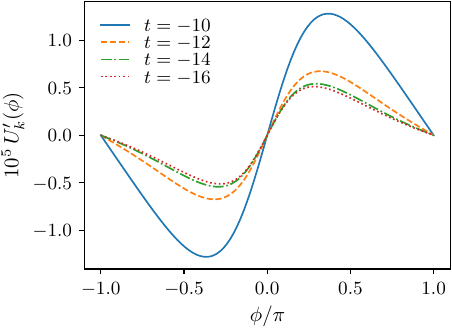}}
	\caption{$U_k'(\phi)=k\tilde U'_k(\phi)$ vs $t=\ln k$ in the insulating (top) and superconducting (bottom) phases. $E_*=\lim_{k\to 0}U''_k(0)$ vanishes in the former case but is finite in the latter as shown in Fig.~\ref{fig_Estar}. [$E_{J}=10^{-2} E_{C}$, $W\to \infty$, $\alpha=0.6$ (top) $\alpha=1.4$ (bottom).]
	}
	\label{fig_Upp} 
\end{figure}

\subsubsection{Mobility and conductance}

At the fixed point, $Z_{2,k}(\phi)=\tilde Z_{2,k}(\phi)/k \to \tilde Z_2^*(\phi)/k$ for $k\to 0$ whereas $U_k(\phi)=k\tilde U_k(\phi)\to 0$. The divergence of $Z_{2,k}(\phi)$ is not a problem since $Z_{2,k}(\phi)\wn^2$ remains finite in the domain of validity of the FE$_2$ ($|\wn|\ll k$). It indicates however that the $\wn^2$ dependence of the self-energy $\Delta_k(i\wn,\phi)$ is not preserved by the RG flow in the regime $k\ll |\wn|$ and one expects $\Delta_{k=0}(i\wn,\phi)\sim|\wn|$ at low energies. Heuristically one can obtain this result by stopping the flow of $Z_{2,k}(\phi)$ at $k\sim |\wn|$ since one expects $\wn$ to play the role of an infrared cutoff when computing the two-point vertex $\Gamma^{(2)}_k(i\wn,\phi)$. Assuming $\Delta_{k=0}(i\wn,0)=C|\wn|$, with $C$ a function of $\alpha$, one then deduces from~(\ref{mob2}) that at the Schmid transition the dc mobility 
\beq  
\mu = \frac{1}{\frac{\alpha}{2\pi} + C}  
\label{mob3} 
\eeq 
takes a nontrivial (i.e. different from 0 and $2\pi/\alpha$) value that depends only on $\alpha$. As a result, the dc conductance $\calG=q^2/\mu$ also takes a nontrivial value. The FE$_2$ does not allow us to determine the value of the $\alpha$-dependent constant $C$ but a more refined approximation scheme, e.g. the Blaizot--Mendez-Galain--Wschebor approximation~\cite{Blaizot06,Benitez09,Benitez12}, would yield the whole frequency dependence of the self-energy (regardless of the value of $|\wn|/k$) and thus the mobility $\mu$ and conductance $\calG$ along the transition line. 
		
\begin{figure}[b]
	\centerline{\includegraphics{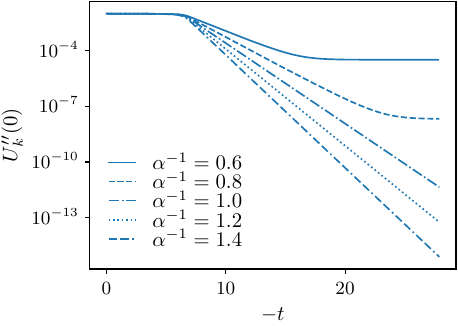}} 
	\centerline{\includegraphics{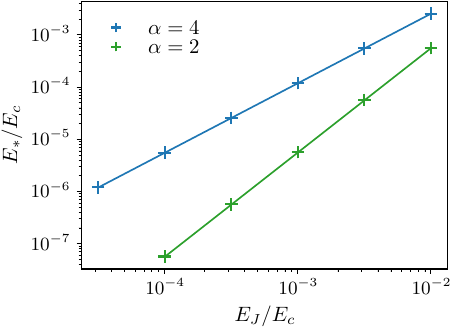}} 
	\caption{(Top) $U''_k(0)$ vs $t=\ln k$ for various values of $\alpha$ and $E_J=0.01$, $1/E_C=0.02$. $E_*=\lim_{k\to 0}U''_k(0)$ takes a nonzero value in the superconducting phase. (Bottom) $E_*$ vs $E_{J}$ in the superconducting phase. The solid lines show the power law behavior $E_* \sim E_J^{\alpha/(\alpha-1)}$.}
	\label{fig_Estar} 
\end{figure}

\section{Superconducting and insulating phases} 

In this section, we compute various observables in order to characterize the insulating and superconducting phases, and compare, when possible, with known results obtained from integrability methods and Monte Carlo simulations. 

\subsection{RG flows} 

Figure~\ref{fig_Upp} shows the flow of the function $U_k'=k\tilde U_k'$ in the insulating and superconducting phases. Although $U'_{k=0}(\phi)=0$, as imposed by the periodicity and convexity of the effective potential~\cite{not11}, the behavior near $\phi=0$ is markedly different in each phase. In the insulating phase, $U'_k(\phi)$ decreases with $k$ and $U_k''(\phi)\to 0$ for all $\phi$. On the other hand, in the superconducting phase $U''_k(\phi)$ reaches a nonzero value in the neighborhood of $\phi=0$. Although this neighborhood shrinks as $k\to 0$, this implies that $E_*=\lim_{k\to 0}U_k''(0)$ takes a nonzero value (Fig.~\ref{fig_Estar}); $U''_k(\phi)$ converges non-uniformally towards $U''_{k=0}(\phi)=0$~\cite{not19}. We conclude that the characteristic energy scale $E_*$ vanishes in the insulating phase but is nonzero in the superconducting phase where it varies as 
\beq 
E_* \sim E_C \left(\frac{E_J}{E_C}\right)^{\alpha/(\alpha-1)} 
\eeq 
in the large bandwidth limit $W\gg E_C$, as shown in Fig.~\ref{fig_Estar}. When approaching the transition, $E_*$ vanishes with the exponent $\nu$, i.e. $E_*\sim (\epj-\epj^{(c)})^{\nu}$ if $\epj$ is varied at fixed $\epc$ (see Fig.~\ref{fig_mob}).  


\begin{figure}[t]
	\centerline{\includegraphics{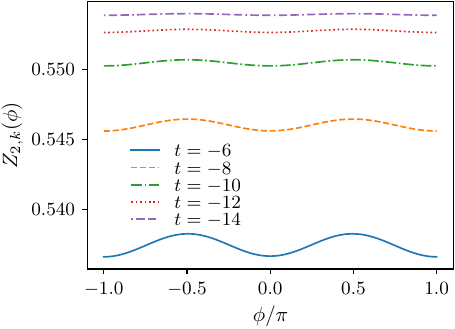}} 
	\centerline{\includegraphics{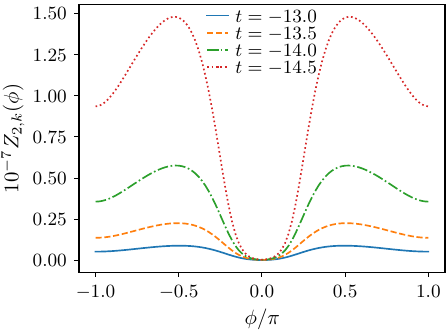}}
	\caption{$Z_{2,k}(\phi)$ vs $t=\ln k$ in the insulating (top) and superconducting (bottom) phases. [$E_{J}=10^{-2}E_{C}$, $W\to\infty$, $\alpha=0.6$ (top) $\alpha=1.4$ (bottom)]}
	\label{fig_Z2} 
\end{figure}

\begin{figure}
\centerline{\includegraphics{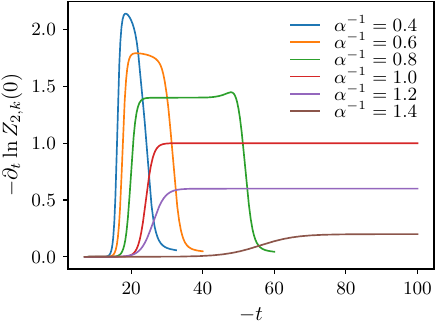}}  \vspace{0.5cm} 
\centerline{\includegraphics{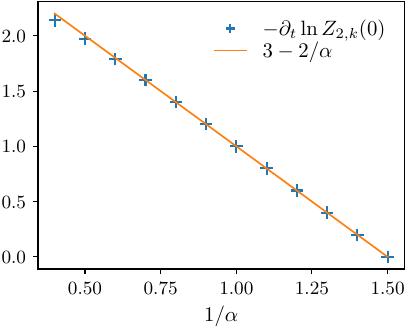}} 
\caption{(Top) $-\dt \ln Z_{2,k}(0)$ vs $t=\ln k$. (Bottom) The value of $-\dt \ln Z_{2,k}(0)$ on the plateau (as shown in the top panel) vs $\alpha$. The system is in the superconducting phase for $\alpha \geq 1$, and in the insulating phase otherwise. [$E_{J}/E_{C}=2\cdot10^{-5}$, $W\to\infty$.]}
\label{fig_eta} 
\end{figure}

The function $Z_{2,k}(\phi)$ is shown in Fig.~\ref{fig_Z2}. While in the insulating phase $Z_{2,k}(\phi)$ becomes constant for $k\to 0$, in the superconducting phase it is strongly non-monotonous, with a form that is reminiscent of the $(1+1)$-dimensional sine-Gordon model~\cite{Daviet19}, and takes large values. At intermediate stages of the flow, when $k\gg E_*$, we find that $-\dt \ln Z_{2,k}(0)\simeq 3-2/\alpha$ takes a $k$-independent value (Fig.~\ref{fig_eta}), which indicates that the self-energy behaves as $\Delta_{k=0}(i\wn)\sim |\wn|^{2/\alpha-1}$ when $|\wn|\gg E_*$. The flow of $Z_{2,k}(\phi)$ however always stops when $k\ll E_*$ and $Z_{2,k}(\phi)$ reaches a nonzero value, implying that the self-energy $\Delta_{k=0}(i\wn,\phi)\sim \wn^2$ is a  quadratic function of the frequency for $\wn\to 0$ (as in the insulating phase). 

These results imply that the propagator can be written in the form 
\beq 
G_{k=0}(i\wn) = \frac{1}{\frac{\alpha}{2\pi}|\wn| + E_*}
\label{propa}
\eeq 
in the low-energy limit (neglecting the $\wn^2$ term). In the insulating phase the propagator $G_{k=0}$ is field independent whereas Eq.~(\ref{propa}) holds for $\phi=0$ in the superconducting phase. Equation~(\ref{propa}) is correct in the insulating phase (where $E_*=0$) and in the superconducting phase when $\alpha\gtrsim 2$; in the latter case the physics is dominated by small fluctuations of the phase about the minima of the periodic potential $-E_J\cos(\varphi)$. On the other hand, Equation~(\ref{propa}) is not correct when $1\leq\alpha\lesssim 2$, a regime where transitions between neighboring minima (instantons) play a crucial role~\cite{Guinea95}.

\subsection{Mobility and admittance}

\begin{figure}
		\includegraphics{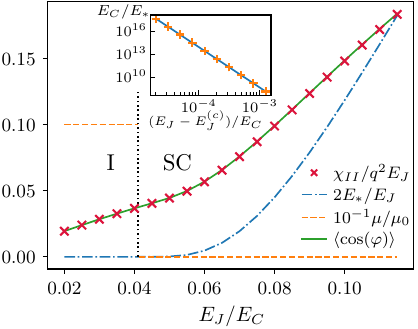}
		\caption{Energy scale $E_*$, dc mobility $\mu=\mu(\w=0)$, coherence $\mean{\cos\varphi}$ and current-current correlation function $\chi^R_{II}(0)$ vs $\epj$ in the large bandwidth limit $W/\epc\to\infty$ for $\alpha\simeq 0.909$. The transition, indicated by the black dotted vertical line, occurs for $\epj^{(c)}/\epc\simeq 0.0432$. The inset shows the divergence of $1/E_*$ in the superconducting phase as the transition is approached. The blue line is a fit $1/E_*\sim (\epj-\epj^{(c)})^{-\nu}$ with the exponent $\nu\simeq 5.3$ obtained from the linearized flow equations (Fig.~\ref{fig_FP}, bottom panel). 
		}
		\label{fig_mob}
\end{figure}

The real part of the mobility $\mu(\w)=\mu'(\w)+i\mu''(\w)$ is deduced from~(\ref{propa}),
\beq 
\mu'(\w) = \frac{\alpha}{2\pi} \frac{\w^2}{\bigl(\frac{\alpha}{2\pi}\w\bigr)^2 + E^2_*} .
\label{mob4}
\eeq 
Thus the dc mobility $\mu(\w=0)$ is equal to $2\pi/\alpha=\mu_0$ in the insulating phase and vanishes in the superconducting phase (Fig.~\ref{fig_mob}). However, the expression of the propagator~(\ref{propa}) being incorrect when $1\leq\alpha\lesssim 2$, the frequency dependence of the mobility~(\ref{mob4}) is not correct in this regime and should vanish as
\beq
\mu'(\w) \sim |\w|^{2\alpha-2}
\label{mob5}
\eeq 
when $\w\to 0$~\cite{Enss05,Meden08,Freyn11}. The behavior of the self-energy, $\Delta_k(i\wn)\sim |\wn|^{2/\alpha-1}$, in the frequency range $|\wn|\gg E_*$ allows us to recover the perturbative (high-frequency) expansion of the mobility $\mu'(\w)=\mu_0-\const\, |\w|^{2/\alpha-2}$~\cite{Kane92}, but the FE$_2$ expansion fails to reproduce the low-energy behavior~(\ref{mob5}). 

From~(\ref{Ydef}), we deduce the low-frequency behavior of the admittance~\cite{not7} 
\beq 
Y(\w) = \llbrace 
\begin{array}{ll} 
	\dfrac{1}{R} & \mbox{  (I)} , \\ 
	\dfrac{1}{R} + \dfrac{i}{L_{\rm eff}(\w+i0^+)} & \mbox{  (SC)} .
\end{array}
\right. 
\label{Y1}
\eeq 
Thus the junction has a vanishing transmission in the insulating phase (I) ---the whole current flowing through the resistor--- and behaves as an effective inductance $L_{\rm eff}=1/q^2E_*$ in the superconducting phase (SC). At the beginning of the flow (i.e. at the classical level), where $U_{\kin}(\phi)=-\epj\cos\phi$, one finds the expected result $1/L_{{\rm eff},\kin}=q^2U''_{\kin}(0)=q^2 \epj$. Note that the inductive response in the superconductive phase follows solely from the vanishing of the self-energy for $\wn\to 0$, i.e. $\Delta_k(i\wn=0,\phi)=0$. The value of $E_*=\lim_{k\to 0}U''_k(0)$, and whether it vanishes or not, however requires to solve the FRG flow equations.

\subsection{Coherence and current-current correlation function}
	
\begin{figure}[t]
	\centerline{\includegraphics{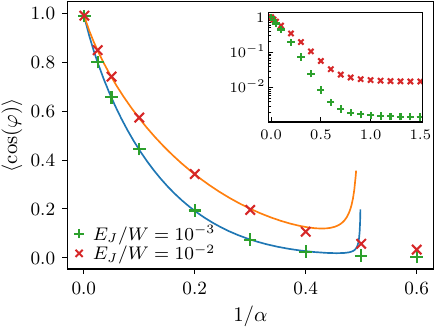}}
	\caption{Expectation value $\mean{\cos\varphi}$ vs $\alpha$ in the superconducting phase when $1/\epc=0$. The lines show the exact result obtained from integrability methods~\cite{Fateev97}.  
	}
	\label{fig_cosexact} 
\end{figure}

The expectation value $\mean{\cos\varphi}$ is computed by adding to the action a time-independent external complex source $h$. The effective potential $U_k(\phi,h^*,h)$ is then a function of $\phi$, $h^*$ and $h$ and $\mean{\cos\varphi}=-U^{(1,0)}(0)$ where $U^{(1,0)}(\phi)=\partial_{h^*} U(\phi,h^*,h)|_{h^*=h=0}$ (see Appendix~\ref{app_coherence} for details). Contrary to the NRG result of Ref.~\cite{Masuki22}, we find that the coherence $\mean{\cos\varphi}$ never vanishes, although it becomes very small in the insulating phase (see the inset in Fig.~\ref{fig_cosexact}), and does not allow one to discriminate between the superconducting and insulating ground states of the RSJJ (Fig.~\ref{fig_mob}). One cannot discard the possibility that a more accurate description of the superconducting phase in the range $1\leq \alpha\lesssim 2$, taking into account the instantons connecting neighboring minima of the periodic potential (see the discussion at the end of the previous section), would give a vanishing coherence in the neighboring of $\alpha=1$, but such a scenario seems rather unlikely. 

If we expand the potential $U^{(1,0)}(\phi)=\sum_{n=0}^\infty u_n^{(1,0)}\cos(n\phi)$ in circular harmonics, we observe that in the insulating phase (where we would naively expect the coherence to vanish since the effective potential is irrelevant), the nonzero value of $U^{(1,0)}(0)$ is entirely due to the zeroth-order harmonic amplitude $u_0^{(1,0)}$ so that the nonzero value of the coherence comes from the dependence of the free energy on the external source $h$. Even though the Josephson coupling is irrelevant in the insulating phase, the coherence cannot be simply computed from the Gaussian action obtained by setting $\epj=0$ (in which case one would find $\mean{\cos\varphi}=0$), we must keep track of the contribution of the high-energy modes to the free energy and its dependence on the external source $h$. 

Because of the U(1) invariance of the action~(\ref{action}), i.e. the invariance in the shift $\varphi(\tau)\to\varphi(\tau)+a$ of the field by an arbitrary constant, the coherence is related to the current-current correlation function $\chi^R_{II}(\w)$ (with $I=q\epj\sin\varphi$ the current through the junction) by~\cite{Safi11}
\beq 
	q^2\epj \mean{\cos\varphi} - \chi^R_{II}(\w=0) = 0 , 
	\label{fsum}
\eeq 
which is the analog of the $f$-sum rule in electron systems with $q^2\epj \mean{\cos\varphi}$ playing the role of the diamagnetic term. This relation is well satisfied by the FRG results (Fig.~\ref{fig_mob}). 

The free energy $f=-\frac{1}{\beta}\ln\calZ$ of the boundary sine-Gordon model is known exactly in the superconducting phase when $\alpha>2$ and $1/\epc=0$~\cite{Fateev97}. Using $\mean{\cos\varphi} = -\partial f/\partial \epj$, one obtains 
\beq 
\mean{\cos\varphi} = \frac{\Gamma\left(\frac{\alpha-2}{2\alpha-2}\right)\Gamma\left(\frac{1}{2\alpha-2}\right)} 
{2\pi^{3/2}\left(1-\frac{1}{\alpha} \right) \frac{\epj}{2bW}} \left( \frac{\pi \epj}{2bW\Gamma(1/\alpha)} \right)^{\alpha/(\alpha-1)} .
\label{cosexact} 
\eeq   
Note that a finite UV cutoff $W$ is necessary to make the boundary sine-Gordon model well defined when $1/\epc=0$. $b$ is a scale factor depending on the implementation of the UV cutoff; for a hard cutoff, $b=e^\gamma/2$ with $\gamma$ the Euler constant~\cite{Daviet19}. Figure~\ref{fig_cosexact} shows that the FRG reproduces the exact result~(\ref{cosexact}) with a very good accuracy. 

\begin{figure} 
	\centerline{\includegraphics{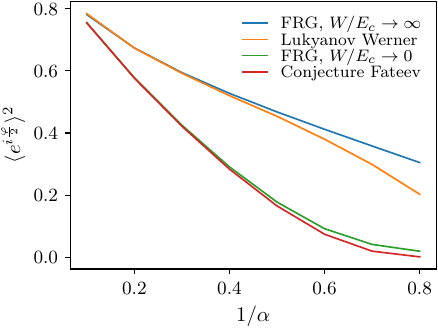}}
	\caption{Expectation value $\mean{\cos(\varphi/2)}$ in the superconducting phase obtained from the FRG  and compared with the conjecture of Fateev {\it et al.}~\cite{Fateev97} in the case $1/E_C=0$, $W<\infty$, and the Monte Carlo simulations of Lukyanov and Werner~\cite{Lukyanov07} in the case $1/E_C>0$, $W\to\infty$. }
\label{fig_LW}
\end{figure}

Finally we discuss the expectation value $\mean{\cos(\varphi/2)}$, whose exact expression has been conjectured by Fateev {\it et al.} in the case $1/E_C=0$ and $W<\infty$ (see Eq.~(3) in Ref.~\cite{Fateev97}). It has also been computed by Lukyanov and Werner in the case $1/E_C>0$ and $W\to\infty$~\cite{Lukyanov07}. Figure~\ref{fig_LW} shows a comparison with the FRG results. In both cases ($1/E_C=0$, $W<\infty$ and $1/E_C>0$, $W\to\infty$), there is a good agreement for $\alpha\gtrsim 2$, which shows that the FE$_2$ is fully (quantitatively) reliable for these values of $\alpha$. However the agreement deteriorates as $\alpha$ approaches one, again a sign that the FE$_2$ is not reliable in the range $1\leq\alpha\lesssim 2$.

\section{Conclusion} 

Our FRG study of the RSJJ has met with mixed success. On the one hand, it clearly shows that the Schmid transition is not controlled by a single fixed point but by a line of critical fixed points, in agreement with the conclusions of Werner, Troyer and Lukyanov based on Monte Carlo simulations and integrability methods~\cite{Werner05,Lukyanov07}. The same conclusion can be drawn from the NRG calculations of Masuki {\it et al.}~\cite{Masuki22}.  

On the other hand there are strong discrepancies regarding the location of the transition line in the plane $(\alpha,\epj/\epc)$. In the large bandwidth limit $W/\epc\gg 1$, Werner and Troyer find a vertical line located at $\alpha=1$ with an accuracy of 2\% (so that the transition can only be induced by varying $\alpha$), the NRG indicates that the transition line is curved in the region $\alpha<1$ in a concave way (except for a surprising reentrance of the superconducting phase at small $\alpha$), whereas the FRG gives a convex transition line which starts with an infinite slope at $\alpha=1$. As already pointed out, the FRG fails in the limit $\alpha\to 0$ and its prediction for the location of the transition line may not be reliable even in the vicinity of $\alpha=1$. These disagreements clearly call for further studies, in particular numerical. The FRG also fails to capture the correct frequency dependence of the mobility in the superconducting phase when $1\leq \alpha\lesssim 2$. 

These two shortcomings ---the spurious phase transition at $\alpha=0$ and the inability to capture the frequency dependence of the mobility when $1\leq \alpha\lesssim 2$--- can be ascribed to the failure of the FRG in describing the instantons between neighboring minima of the periodic potential. This is in sharp contrast with the $(1+1)$ sine-Gordon model where a derivative expansion of the effective action to second order is sufficient to compute the mass of the solitons and the lowest-lying breather (soliton-antisoliton bound state) with high accuracy~\cite{Daviet19}. The description of topological defects is an open issue in the FRG approach. Aside from the topological excitations of the $(1+1)$-dimensional sine-Gordon model, the FRG provides us with a good description of the vortices and the Berezinskii-Kosterlitz-Thouless transition in the two-dimensional O(2) model~\cite{Graeter95,Gersdorff01,Jakubczyk14}, but a good description of the kinks in the one-dimensional Ising has still not been achieved~\cite{Rulquin16,Farkas23}. Whether or not the FRG approach, used here in combination with a second-order frequency expansion, can be improved in order to describe the instantons of the boundary sine-Gordon model is an open issue. 

The discrepancies between the various theoretical and numerical approaches could in principle be settled by future experiments. By determining the location of the transition line one could distinguish between the two scenarios proposed in Fig.~\ref{fig_phasediag}. We have provided a detailed prediction of this location near $\alpha=1$, as a function of $W$ and $E_C$, in the scenario where the transition line bends in the region $\alpha<1$ (top panel in Fig.~\ref{fig_phasediag}), which could be checked in finite-frequency measurements of a RSJJ~\cite{Murani20,Kuzmin23}.

\section*{Acknowledgment}
We thank C. Altimiras, P. Azaria, D. Esteve, S. Florens, P. Joyez, P. Lecheminant, H. le Sueur, C. Mora, I. Safi, H. Saleur, G. Takacs and W. Zwerger for discussions and/or correspondence.

\widetext	

\appendix

\section{Flow equations} 
\label{app_floweq} 

The flow equations of $U_k(\phi)$ and $Z_{2,k}(\phi)$ are obtained by relating these two quantities to the two-point vertex
\beq 
\Gamma_k^{(2)}(i\wn,\phi) = Z_1 |\wn| + Z_{2,k}(\phi)\wn^2 + U_k''(\phi) 
\eeq 
in a time-independent field $\phi(\tau)=\phi$. Thus we have 
\beq 
\begin{split}
	\dt U_k''(\phi) &= \dt \Gamma_k^{(2)}(0,\phi) , \\
	\dt Z_{2,k}(\phi) &= \dt \frac{\Gamma_k^{(2)}(i\w_1,\phi) - \Gamma_k(0,\phi)}{\w_1^2} 
\end{split}
\label{app4}
\eeq 
where $\omega_1=2\pi T$ and we use a discrete frequency derivative for $Z_{2,k}(\phi)$. $\Gamma_k^{(2)}$ satisfies the flow equation 
\begin{align}
	\dt \Gamma_k^{(2)}(i\wnu,\phi) ={}& \sum_{\wn} \dt R_k(i\wn) \biggl\{ - \half G_k(i\wn,\phi)^2 \Gamma_k^{(4)}(i\wnu,-i\wnu,i\wn,-i\wn,\phi) \nonumber\\ & + G_k(i\wn,\phi)^2 \Gamma_k^{(3)}(i\wnu,i\wn,-i\w_{n+\nu},\phi) G_k(i\w_{n+\nu},\phi) \Gamma_k^{(3)}(-i\wnu,i\w_{n+\nu},-i\wn,\phi)  
	\biggr\} ,
	\label{app5} 
\end{align}
where all quantities are evaluated in a time-independent field $\phi$. In the limit $T\to 0$, the Matsubara frequency becomes a continuous variable and the discrete sums can be replaced by integrals. Likewise the discrete derivative in~(\ref{app4}) becomes a standard derivative. However, the propagator $G_k(i\wn)$ being a non-analytic function of $\wn$, care must be taken when taking the derivative wrt $\wnu^2$ of the rhs of~(\ref{app5}) since a naive expansion in $\wnu$ of the integrand may give wrong results~\cite{Balog14,not15}. 

The flow equations for the dimensionless functions defined in~(\ref{dimless}) read
\begin{align}
	\dt \tilde U''_k ={}& -\tilde U''_k + 2 l_2^3 \tilde U_k''' \tilde Z_{2,k}' + l_0^3 \tilde U'''_k{}^2 - \half l_0^2 \tilde U^{(4)}_k +  l_4^3 \tilde Z_{2,k}'{}^2 - \half l_2^2 \tilde Z_{2,k}'' , \label{app1} \\ 
	\dt \tilde Z_{2,k} ={}&  \tilde Z_{2,k} +  \frac{1}{2 \tilde\omega_1^2} \Bigl\{ 2 l_0^{2,1}(i\tilde\omega_1 ) [ \tilde U_k''' + \tilde\omega_1 ^2 \tilde Z_{2,k}' ]^2 + 4 \tilde\omega_1  l_1^{2,1}(i\tilde\omega_1 )
	\tilde U_k''' \tilde Z_{2,k}'+4 l_2^{2,1}(i\tilde\omega_1 ) \tilde U_k''' \tilde Z_{2,k}' - 4 l_2^3
	\tilde U_k''' \tilde Z_{2,k}'  - 2 l_0^3 \tilde U_k'''{}^2 \nonumber \\ 
	& \hspace{-0.2cm} +4 \tilde\omega_1 ^3 l_1^{2,1}(i\tilde\omega_1 )
	\tilde Z_{2,k}'{}^2+6 \tilde\omega_1 ^2 l_2^{2,1}(i\tilde\omega_1 ) \tilde Z_{2,k}'{}^2-l_0^2 \tilde\omega_1 ^2
	\tilde Z_{2,k}''  + 4 \tilde\omega_1  l_3^{2,1}(i\tilde\omega_1 ) \tilde Z_{2,k}'{}^2+2 l_4^{2,1}(i\tilde\omega_1 )
	\tilde Z_{2,k}'{}^2 - 2 l_4^3 \tilde Z_{2,k}'{}^2 \Bigr\} , \label{app2} 
\end{align}
where $\tilde\omega_n=\wn/k=2\pi n\tilde T$ with $\tilde T=1/\tilde \beta=T/k$ the dimensionless temperature. The prime, double prime, etc., denote derivatives wrt $\phi$. We have introduced the threshold functions 
\beq 
\begin{split}
	l_p^m ={}& \frac{1}{\tilde\beta} \sum_{\twn} \dot {\tilde R}_k(i\twn) \tilde G_k(i\twn,\phi)^m |\twn|^p , \\ 
	l_p^{m_1,m_2}(i\twnu) ={}& \frac{1}{\tilde\beta} \sum_{\twn} \dot {\tilde R}_k(i\twn) \tilde G_k(i\twn,\phi)^{m_1}  \tilde G_k(i\tilde\w_{n+\nu},\phi)^{m_2} \twn^p 
\end{split} 
\label{thres} 
\eeq  
and the dimensionless cutoff function and its time derivative,
\begin{align} 
	\tilde R_k(i\twn) &= \frac{R_k(i\wn)}{k}= \frac{\alpha}{2\pi} |\twn| r(|\twn|) + kZ_{2,k} \twn^2 r(\twn^2) , \\ 
	\dot{\tilde R}_k(i\twn) &= \frac{\dt R_k(i\wn)}{k} = - \frac{\alpha}{2\pi} \twn^2 r'(|\twn|) - 2 kZ_{2,k} \twn^4 r'(\twn^2) + k\dt Z_{2,k} \twn^2 r(\twn^2)  , 
\end{align} 
where
\begin{equation}
	\tilde G_k(i\twn,\phi) = k G_k(i\wn,\phi) 
	= \frac{1}{\frac{\alpha}{2\pi}|\twn| + \tilde Z_2(\phi) \twn^2 + \tilde U_k''(\phi) + \tilde R_k(i\twn)} 
\end{equation} 
is the dimensionless propagator in a time-independent field $\phi(\tau)=\phi$. In the zero-temperature limit, the Matsubara sums in~(\ref{thres}) become integrals over the continuous variable $\tw$, 
\beq 
\frac{1}{\tilde\beta} \sum_{\twn} \to \intinf \frac{\tw}{2\pi} \qquad (T\to 0). 
\eeq 
To alleviate the notations, we do not write explicitly the dependence of the threshold functions on $k$,  $\tilde Z_{2,k}(\phi)$ and $\tilde U''_k(\phi)$. 

For $\tilde Z_{2,k}(\phi)=\tilde U_k''(\phi)=0$ and in the limit $T\to 0$, 
\begin{equation} 
	\bar l_0^2 = 
	l_0^2|_{\tilde Z_{2,k}=\tilde U''_k=0} 
	= \intinf \frac{d\tw}{2\pi} \dot{\tilde R}_k(i\tw) \tilde G_k(i\tw)^2 
	= - \frac{2}{\alpha} \int_0^\infty d\tw \frac{r'(\tw)}{[1+r(\tw)]^2} 
	= \frac{2}{\alpha} .
\end{equation} 
The threshold function $\bar l_0^2$ is universal, i.e. independent of the cutoff function $R_k$, provided that  $r(0)=\infty$ and $r(\infty)=0$.

\section{Perturbation theory from truncated flow equations} 
\label{app_perturbation} 

We use the flow equations to reconstruct the perturbation theory near $\alpha=1$ where the critical point is close to the Gaussian fixed point. In a first step, we approximate $U_k(\phi)$ and $Z_{2,k}(\phi)$ by retaining only the coupling constants that are nonzero for the initial condition at $k=\kin$, i.e. 
\beq 
U_k(\phi) = - \epjk \cos\phi , \qquad  
Z_{2,k}(\phi) = \frac{1}{2\epck} .
\label{UZ2trunc}
\eeq 
Anticipating that the fixed point values $1/\tepc^*$ and $\tepj^*$ of the dimensionless coupling constants $1/\tepck=k/\epck$ and $\tepjk=\epjk/k$ are of order $\eps=1/\alpha-1$ and $\sqrt{\eps}$, respectively, we include in the flow equations all terms up to order $\eps^{3/2}$. To do so, one must expand the threshold functions as follows, 
\beq 
l_p^m = \bar l_p^m - m \bigl[ \tilde U_k''(\phi) \bar l_p^{m+1} + \tilde Z_{2,k}(\phi) \bar l_{p+2}^{m+1} \bigr] + \half (m^2+m)  \tilde U_k''(\phi)^2 \bar l_p^{m+2} ,
\label{lexp1}
\eeq  
where $\bar l_p^m=l_p^m|_{\tilde U''_k=\tilde Z_{2,k}=0}$. For simplicity, we consider here the cutoff function $R_k(i\twn)=(\alpha/2\pi)|\twn|r(|\twn|)$. This yields the flow equations 
\beq 
\begin{split} 
	k \dk \frac{1}{\tepck} &= \frac{1}{\tepck} + \blp \tepjk^2 , \\ 
	k \dk \tepjk &= \tepjk\left(-1 + \frac{1}{\alpha} \right) - \half \bl \frac{\tepjk}{\tepck} + \frac{3}{8} \bar l_0^4 \tepjk^3,
\end{split}
\label{rgeqtrunc} 
\eeq 
where
\begin{equation}
	\bar l_p^{m_1,m_2}{}' = 
	\llbrace  \begin{array}{ll}	
		\lim_{\tw\to 0} \dfrac{\bar l_p^{m_1,m_2}(i\tw) - \bar l_p^{m_1+m_2}}{\tw^2}
		= \partial_{\tw^2} \bar l_p^{m_1,m_2}(i\tw) \Bigl|_{\tw=0} & (p \mbox{ even}) , \\	
		\lim_{\tw\to 0} \dfrac{\bar l_p^{m_1,m_2}(i\tw)}{\tw}
		= \partial_{\tw} \bar l_p^{m_1,m_2}(i\tw) \Bigl|_{\tw=0} & (p \mbox{ odd}) ,
	\end{array} 
	\right.  
\end{equation}
using $\bar l_p^{m_1,m_2}(i\tw=0)=\bar l_p^{m_1+m_2}$ (0) for $p$ even (odd). We denote by $\bar l_p^{m_1,m_2}(i\tw)$ the function $l_p^{m_1,m_2}(i\tw)$ in the limit $\tilde U_k''=\tilde Z_{2,k}=0$. The parameters $\bl>0$, $\bar l_0^4>0$ and $\blp<0$ in~(\ref{rgeqtrunc}) are real numbers whose values depend on the function $r$ discriminating between low ($|\wn|\lesssim k$) and high ($|\wn|\gtrsim k$) frequency modes. 

When $\alpha\to 1$, the running of the variable $1/\tepck$ is initially much faster than that of $\tepjk$; after a transient regime the value of $1/\tepck$ is entirely determined by the value of $\tepjk$, 
\beq 
\frac{1}{\tepck} = - \blp \tepjk^2 . 
\label{pt1} 
\eeq 
In other words, all RG trajectories in the $(1/\tepck,\tepjk)$ plane collapse on a single line as shown in Figs.~\ref{fig_flowdiag1} and \ref{fig_flowdiag2}: For a general discussion of this ``large-river effect'', see Refs.~\cite{Bagnuls01,Bagnuls01a}. The flow equation on that line is deduced from~(\ref{rgeqtrunc}) and (\ref{pt1}),
\beq 
k\dk \tepjk = \tepjk\left(-1 + \frac{1}{\alpha} \right)  + \left( \frac{3}{8} \bar l_0^4 + \half \bl\blp \right) \tepjk^3 . 
\label{pt2}
\eeq 
We thus obtain the nontrivial fixed point  
\beq 
\tepj^* = \biggl( \frac{8(\alpha-1)/\alpha}{4 \bl\blp + 3 \bar l_0^4} \biggr)^{1/2} , \qquad 
\frac{1}{\tepc^*} =  -\frac{8\blp (\alpha-1)/\alpha}{4 \bl\blp + 3 \bar l_0^4}.
\label{pt3}
\eeq 
Depending on the sign of $4 \bl\blp + 3 \bar l_0^4$, this fixed point exists for $\alpha>1$ or $\alpha<1$. 
Linearizing the flow equation~(\ref{pt2}) yields the critical exponent 
\beq 
\frac{1}{\nu} = 2 \left( \frac{1}{\alpha} - 1 \right) \quad \mbox{for} \quad \alpha\to 1 . 
\label{pt4}
\eeq 
The same result can be obtained by linearizing Eqs.~(\ref{rgeqtrunc}) about the fixed point~(\ref{pt3}). 

To ensure that Eq.~(\ref{pt4}) is correct one should also consider the coupling constants that are not included in the ansatz~(\ref{UZ2trunc}). We thus use the harmonic expansion
\beq 
\tilde U_k(\phi) = \sum_{n=0}^\infty \tilde u_{n,k} \cos(n\phi) , \qquad 
\tilde Z_{2k}(\phi) = \sum_{n=0}^\infty \tilde z_{n,k} \cos(n\phi)
\eeq  
and expand the flow equations in powers of $\eps=1/\alpha-1$. In addition to~(\ref{lexp1}) one must expand the threshold function 
\begin{align}
	l_p^{m_1,m_2}(i\twnu) ={}& \bar l_p^{m_1,m_2}(i\twnu) - m_1 \bigl[ \tilde U_k''(\phi) \bar l_p^{m_1+1,m_2}(i\twnu) + \tilde Z_{2,k}(\phi) \bar l_{p+2}^{m_1+1,m_2}(i\twnu) \bigr] + \half (m_1^2+m_1)  \tilde U_k''(\phi)^2 \bar l_p^{m_1+2,m_2}(i\twnu) \nonumber\\ & 
	- m_2 \bigl[ \tilde U_k''(\phi) \bar l_p^{m_1,m_2+1}(i\twnu) + \tilde Z_{2,k}(\phi) \bar l_{p+2}^{m_1,m_2+1}(i\twnu) \bigr] + \half (m_2^2+m_2)  \tilde U_k''(\phi)^2 \bar l_p^{m_1,m_2+2}(i\twnu) \nonumber\\ & + m_1 m_2 \tilde U_k''(\phi)^2 \bar l_p^{m_1+1,m_2+1}(i\twnu) . 
\end{align}
Near the fixed point, $\tilde u_{1,k}=\calO(\sqrt{\eps})$, $\tilde z_{0,k},\tilde z_{2,k},\tilde u_{n>1,k}=\calO(\eps)$ and $\tilde z_{n\neq 0,2,k}=\calO(\eps^{3/2})$. Using the fact that the running of $\tilde u_{1,k}$ is initially much slower than the other variables, after a transient regime we find 
\beq  
\begin{split}
	\tilde z_{0,k} &= - \half \bar l_0^{2,1}{}'  \tilde u^2_{1,k} , \\ 
	\tilde z_{1,k} &= - \frac{\alpha}{4(1+\alpha)} 
	\bigl[ 32 \bar l_0^{2,1}{}' \tilde u_{2,k} +  \bigl(\bar l_0^{2,2}{}' + 2 \bar l_0^{3,1}{}' \bigr) \tilde u_{1,k}^2 - 8 \bigl( \bar l_1^{2,1}{}' + \bar l_2^{2,1}{}' \bigr) \tilde z_{2,k} \bigr]  \tilde u_{1,k} , \\ 
	\tilde z_{2,k} &= \frac{\alpha}{2(4+\alpha)}  \bar l_0^{2,1}{}' \tilde u^2_{1,k} , \\ 
	\tilde u_{2,k} &= \frac{\alpha}{4(\alpha-4)} \bigl( \bar l_0^3 \tilde u_{1,k}^2  - 2\bar l_2^2  \tilde z_{2,k} \bigr) 
\end{split}
\label{pt5}
\eeq  
to leading order in $\eps$, and 
\beq 
\dt \tilde u_{1,k} = \left( \frac{1}{\alpha} - 1 \right) \tilde u_{1,k} + \frac{1}{8} 
\bigl\{ 3 \bar l_0^4 \tilde u_{1,k}^3 + 2 [ 8 \bar l_0^3 \tilde u_{2,k} - 2 \bar l_2^3 (2\tilde z_{0,k}+\tilde z_{2,k}) ] \tilde u_{1,k}
-4 \bar l_2^2 \tilde z_{1,k} \bigr\} 
\label{pt6}
\eeq   
including all terms of order $\eps^{3/2}$. Equations~(\ref{pt5}) and (\ref{pt6}) lead to~(\ref{pt7}), 
where $\calF$ is a complicated combination of the threshold functions $\bar l_p^m$ and $\bar l_p^{m_1,m_2}{}'$.

\section{Current-current correlation function and coherence}
\label{app_coherence} 

To compute the expectation value $\mean{\cos\varphi}$ and the zero-frequency limit of the current-current correlation function $\chi_{II}(i\wn)$, one must introduce a time-independent external complex source $h$ in the action~(\ref{action}), i.e. consider 
\beq 
S - \inttau \bigl(h^* e^{i\varphi(\tau)} + h e^{-i\varphi(\tau)} \bigr) .
\label{action3}
\eeq 
In the FE$_2$ the effective action takes the form~(\ref{Gamma1}) where however the functions $Z_{2,k}(\phi,h^*,h)$ and $U_k(\phi,h^*,h)$ depend on $h^*$ and $h$. We can now use 
\beq 
\begin{gathered}
	\mean{\cos\varphi} = \frac{1}{\beta} \frac{\partial \ln \calZ(h^*,h)}{\partial h} \biggl|_{h^*=h=0} , \\ 
	\chi_{II}(i\wn=0) = - \frac{q^2\epj^2}{4\beta}  \left(  
	\frac{\partial^2}{\partial h^*{}^2} + \frac{\partial^2}{\partial h^2}
	- 2 \frac{\partial^2}{\partial h^* \partial h} \right)  \ln\calZ(h^*,h) \Bigl|_{h^*=h=0} ,
\end{gathered}
\eeq 
where $\calZ(h^*,h)$ is the partition function obtained from~(\ref{action3}). These equations can be rewritten in terms of the effective potential $U(\phi,h^*,h) \equiv U_{k=0}(\phi,h^*,h)$ and $G(i\omega_{n},\phi)\equiv G_{k=0}(i\omega_{n},\phi)$~\cite{not4},  
\begin{align} 
	\mean{\cos\varphi} ={}& - U^{(1,0)}(0) , \\ 
	\chi_{II}(i\wn=0) ={}& - \frac{q^4\epj^2}{4} \Bigl\{ -U^{(2,0)}(0) -U^{(0,2)}(0) + 2 U^{(1,1)}(0) \nonumber\\ & + G(0,0) \bigl[ U^{(1,0)}{}'(0)^2 + U^{(0,1)}{}'(0)^2  - 2 U^{(1,0)}{}'(0) U^{(0,1)}{}'(0)
	\bigr] \Bigr\} ,
\end{align} 
where we use the notation $U^{(i,j)}(\phi)=\partial^i_{h^*} \partial^j_{h} U(\phi)|_{h^*=h=0}$ and the prime denotes a derivation with respect to $\phi$. $U^{(i,j)}(\phi)$ can be obtained from the flow equations of $U_k^{(i,j)}(\phi)$ (which we do not show here). 

A similar method can be used to obtain the expectation value $\mean{\cos(\varphi/2)}$ as well as $\chi(i\wn=0)$ where $\chi(\tau)=\mean{e^{\frac{i}{2}\varphi(\tau)} e^{-\frac{i}{2}\varphi(0)}}$. 



\begin{thebibliography}{61}
\bibitem [{not({\natexlab{a}})}]{not22}%
  \BibitemOpen
  \href@noop {} {} \bibinfo {note} {Such a model is
  also referred to as the resistively and capacitively shunted junction (RCSJ)
  model.}\BibitemShut {Stop}%
\bibitem [{not({\natexlab{a}})}]{not16a}%
  \BibitemOpen
  \href@noop {} {} \bibinfo {note} {Note that in
  Ref.~\cite{Masuki22}, the boundary sine-Gordon model is understood as the
  model defined by the action~(\ref{action}) with $1/\epc=0$.}\BibitemShut
  {Stop}%
\bibitem [{not({\natexlab{b}})}]{not12}%
  \BibitemOpen
  \href@noop {} {} \bibinfo {note} {For the Luttinger liquid
  in the presence of an impurity, the finite value of $1/\epc$ is due to the
  finite range of the impurity potential.}\BibitemShut {Stop}%
\bibitem [{not({\natexlab{c}})}]{not9}%
  \BibitemOpen
  \href@noop {} {} \bibinfo {note} {The frequency expansion
  in~(\ref{Gamma1}) is not a derivative expansion since the Fourier transform
  of $|\wn|$ ($\sim -1/\tau^2$) is nonlocal in time.}\BibitemShut {Stop}%
\bibitem [{not({\natexlab{d}})}]{not13}%
  \BibitemOpen
  \href@noop {} {} \bibinfo {note} {Although we initiate the
  flow at $\kin\gg\min(W,\epc)$, the renormalization of $1/\epck,$ and $\epjk$
  is weak when $k$ varies between $\kin$ and $W$ (when $W\gg\epc$) and
  $1/\tilde E_{C,W}\simeq W/E_C$, $\tilde E_{J,W}\simeq E_J/W$.}\BibitemShut
  {Stop}%
\bibitem [{not({\natexlab{e}})}]{not18}%
  \BibitemOpen
  \href@noop {} {} \bibinfo {note} {We consider $\tilde
  U'_k$ rather than $\tilde U_k$ since the field-independent part of $\tilde
  U_k$ is always relevant.}\BibitemShut {Stop}%
\bibitem [{not({\natexlab{f}})}]{notad}%
  \BibitemOpen
  \href@noop {} {} \bibinfo {note} {Note that this does
  not imply that a non-dissipative JJ is insulating. In the absence of
  dissipation, a JJ must be described by a compact phase variable $\varphi\in
  [0,2\pi]$ and the ground state of the model defined by the
  action~(\ref{action}) is then superconducting in agreement with experimental
  observations. Only in the presence of dissipation does the phase variable
  decompactify~\cite{Morel21}, thus justifying the model considered in the
  manuscript when $\alpha>0$.}\BibitemShut {Stop}%
\bibitem [{not({\natexlab{f}})}]{not14}%
  \BibitemOpen
  \href@noop {} {} \bibinfo {note} {Notice that Werner and
  Troyer's definition of $\epc$ differs from ours by a factor $1/4$: $\epc^{\rm
  WT}=\epc/4$.}\BibitemShut {Stop}%
\bibitem [{not({\natexlab{g}})}]{not11}%
  \BibitemOpen
  \href@noop {} {} \bibinfo {note} {Since the effective
  action $\Gamma_k[\phi]$ is not the Legendre transform of the free energy
  $\ln\calZ_k[J]$ when $k>0$ (due to the subtraction of $\Delta S_k[\phi]$
  in~(\ref{Gammadef}), the effective potential $U_k(\phi)$ must be periodic
  but, unless $k=0$, is not necessary convex.}\BibitemShut {Stop}%
\bibitem [{not({\natexlab{h}})}]{not19}%
  \BibitemOpen
  \href@noop {} {} \bibinfo {note} {In practice it is
  difficult to obtain the convexity of the potential, i.e., for all $\phi$, the
  vanishing of $U'_k(\phi)$ as $k\to 0$.}\BibitemShut {Stop}%
\bibitem [{not({\natexlab{i}})}]{not7}%
  \BibitemOpen
  \href@noop {} {} \bibinfo {note} {The fact that an
  inductance $L$ has the admittance $i/L\w$, rather than $-i/L\w$, as with the
  usual electrical engineering convention, follows from our convention for
  Fourier transforms.}\BibitemShut {Stop}%
\bibitem [{not({\natexlab{j}})}]{not15}%
  \BibitemOpen
  \href@noop {} {} \bibinfo {note} {The propagator
  $G_k(i\w)$ being singular at $\w=0$ (in the zero-temperature limit where
  $\wn$ becomes a continuous variable), one must split the frequency integral
  between different intervals where the integrand is regular before
  expanding in powers of the external frequency $\wnu$.}\BibitemShut {Stop}%
\end{thebibliography}

%

\end{document}